# Study and evaluation of the Ronen Method accuracy at material interfaces

Johan Cufe[a], Daniele Tomatis[b], Erez Gilad[a,*]

[a]*The Unit of Nuclear Engineering,
Ben-Gurion University of the Negev, Beer Sheva 8410501, Israel*
[b]*newcleo SrL, Via Giuseppe Galliano 27, 10129 Turin, Italy*

**Abstract**

The Ronen method (RM) demands for successive resolutions of the diffusion equation where local diffusion constants are modified to reproduce more accurate estimates of the currents by a transport operator. The methodology is currently formulated by using the formalism of the collision probability method (CPM) for the current evaluation and RM was recently tested on a complete suite of one-dimensional multigroup benchmark problems. Small differences in the flux (less than 2%) were reported at material interfaces and close to the vacuum boundary with respect to the reference solution from transport (CPM). In this work, a verification check is first set to prove an equivalence between diffusion and transport when optimal diffusion coefficients are computed by the transport solution itself and employed in a standard diffusion calculation. 1G and 2G criticality problems from the same criticality benchmark test suite of previous publications are tested. Then, the accuracy of the flux distribution near the vacuum boundary and material interfaces is computed using the RM for different approximations of the vacuum boundary and with respect to decreasing values of the RM convergence criterion set in its iterative scheme. Indeed, the RM calculates more accurate flux distribution at all material interfaces, regardless of the initial values used for the diffusion coefficient and the extrapolated distance at the beginning of the iterative process. Maximal flux deviations fall everywhere around 0.01% when the RM convergence criterion is set to ten significant digits, leading to two orders of magnitude improvement in the flux deviation.

## Contents

---

[*]Corresponding author. *Email address:* `gilade@bgu.ac.il`

## 1. Introduction

Full-core calculations aim at obtaining accurate quantities like the neutron flux and reaction rates over the complete reactor. These calculations consist of solving the diffusion equation or some other second-order approximations of transport using a few energy groups G. Two energy groups are sufficient for thermal reactors, while more are necessary for fast reactors. The actual trend in the design of nuclear reactors shows heterogeneous loading patterns where the transport effects are more evident, such that classic diffusion finds quickly its shortcomings to provide reliable results. Indeed, diffusion theory makes the assumptions of smooth flux variations, small absorption compared to scattering and low scattering anisotropy which becomes all questionable for modern core configurations.

The limitations implied by diffusion can be remedied by introducing transport corrections though, as done by the Ronen method (RM), which belongs to the class of transport approximation methods employing diffusion solvers [4, 12]. In the RM, an integral transport equation expresses the net current used in an iterative and nonlinear scheme to force the solution from diffusion to fulfil the same integral equation. The RM was recently tested on a complete suite of one-dimensional multigroup benchmark problems [5, 12]. Although an excellent agreement was always obtained in eigenvalue problems on integral quantities like the neutron multiplication factor $k_{\text{eff}}$, differences in the flux distribution were observed with respect to the reference solution from transport at positions where transport effects are more pronounced, namely near vacuum and material interfaces.

Further investigations on the method are needed to determine whether the RM can reproduce the transport solution with sufficient accuracy at every spatial position. In this work, we first determine optimal values for the diffusion coefficient and the extrapolated distance, aiming to check and enforce an equivalence between diffusion and transport everywhere in the domain. Tests are carried out in 1G-2G homogeneous and 1G heterogeneous slab problems from Sood's critical benchmark test suite [9]. In addition, for the homogeneous case, analytical expressions for the diffusion coefficient and the extrapolated distance are derived by solving the neutron transport equation using Case's method, following the approach by Mitsis [7]. Standard diffusion is tested using these coefficients as input data. Once this check is verified, the RM accuracy is investigated with decreasing convergence tolerance set in its iterative loop, providing an update on RM accuracy estimation.

The theoretical background of the RM is presented in Section 2. The equivalence procedure between diffusion and transport is described in Section 3. The status of the RM performance is summarized in Section 4. The results section including the equivalence check and RM accuracy evaluation is given in Section 5. The article ends with the conclusion in Section 6.

## 2. Theoretical background

In the RM, successive solutions of the diffusion equation are performed with local corrections to the diffusion coefficients that are introduced to reproduce new estimates of the currents obtained by an integral transport operator [12]. The neutron source, used in the integral expression for the current, is calculated with the scalar flux obtained by diffusion. Convergence on the scalar flux and net current distributions is sought through non-linear iterations alternating the diffusive solver and the evaluation of the integral expression. Two main different implementations are currently available, depending on how the current correction is implemented in the streaming term of the neutron balance equation. One is based on the drift-like current while the other redefines *online* the diffusion coefficient using Fick's law. What mainly differs between these two implementations is the way physics is reproduced by the local corrections in the diffusive solver.

In 1D slab geometry with standard finite differences formalism, using integer and rational subscripts for cell-averaged and interface quantities, the neutron current using Fick's law is expressed as

$$J^D_{g,i+1/2} \cong -2D_{g,i+1/2} \frac{\Phi_{g,i+1} - \Phi_{g,i}}{\Delta_{i+1} + \Delta_i}, \quad D_{g,i+1/2} = \frac{\Delta_i + \Delta_{i+1}}{\Delta_i/D_{g,i} + \Delta_{i+1}/D_{g,i+1}} \quad (1)$$

where $\Delta_i = (x_{i+1/2} - x_{i-1/2})$ and the diffusion coefficient is approximated by first-order Taylor expansions and $D_{g,i} = 1/3\sigma_{g,i}$. The spatially discretized integral neutron current with vacuum boundary conditions, accounting also linearly anisotropic scattering sources, is formulated as

$$J_{g,i+1/2} = \frac{1}{2} \sum_{j=0}^{I-1} \frac{q_{0,g,j}}{\sigma_{g,j}} \left( E_3 \left[ \tau_g(x_{j+1/2}, x_{i+1/2}) \right] - E_3 \left[ \tau_g(x_{j-1/2}, x_{i+1/2}) \right] \right) + \\ \text{Sgn}(i-j) + \frac{3}{2} \sum_{j=0}^{I-1} \frac{q_{1,g,j}}{\sigma_{g,j}} \left( E_4 \left[ \tau_g(x_{j+1/2}, x_{i+1/2}) \right] - E_4 \left[ \tau_g(x_{j-1/2}, x_{i+1/2}) \right] \right) \quad (2)$$

with the introduction of the spatial dependent integral exponential functions $E_n$, function of the optical path length $\tau_g$, for order $n = 3, 4$ [5]. The sources $q_{0,g,j}$ and $q_{1,g,j}$ indicate the isotropic and linearly anisotropic scattering sources, respectively defined as

$$q_{0,g,j} = \sum_{g'=1}^{G} (\sigma_{s_0,g' \to g,j} + \frac{\chi_g}{k_{\text{eff}}} \nu \sigma_{f,g',j}) \Phi_{g',j}, \quad q_{1,g,j} = \sum_{g'=1}^{G} \sigma_{s_1,g' \to g,j} J_{g',j} \quad (3)$$

with the current $J_{g',j}$ appearing in the linearly anisotropic scattering source as averaged in the cell $j$. The sums involved in Eq. (2), keeping the isotropic scattering sources only, can be reduced by half if we consider only contributions coming from $i > j$ with the others being just opposite in sign. Thus, Eq. (2) can be also reformulated for the partial currents. Assuming vacuum boundary conditions, we can define the following

$$J^{\pm}_{g,i+1/2} = \sum_{j=1}^{I} q_{0,g,j} \Delta_i \tilde{e}^{\pm}_{g,i+1/2,j} \quad (4)$$

which makes explicit use of the theory of the collision probability method [6], where $\tilde{e}$ represents the escape probability. In case of reflective boundary conditions, the term $J^{\pm}_{g,-1/2}\tilde{t}_{g,i+1/2,j}$ must be added to Eq. (4), where we made use of the transmission probability $\tilde{t}$. Details of this formulation can be found in [12]. The expression of Eq. (2) is used to correct the "diffusive" current of Eq. (1), leading to a more accurate current estimation.

*2.1. Iterative redefinition of the diffusion coefficient*

This implementation was originally suggested by Ronen [8]. The diffusion coefficient is redefined according to Fick's law, taking the current from the integral expression and the scalar flux used for the source at its integrand

$$\tilde{D}_{g,i+1/2} = -\frac{J_{g,i+1/2}}{2\frac{\Phi_{g,i+1}-\Phi_{g,i}}{\Delta x_{i+1}+\Delta x_i}}. \tag{5}$$

The corrected diffusion coefficient can be recast as $\tilde{D}_{g,i+1/2} = D_{g,i+1/2} + \delta D_{g,i+1/2}$ with $D_{g,i+1/2}$ obtained by Eq. (1). The term $\delta D_{g,i+1/2}$ accounts for the correction to the diffusion coefficient. Eq. (5) can show indeterminate division by zero in case of flat flux, requiring proper numerical fix-up [5]. In the RM, a new generalized eigenvalue problem arises at each r-th iteration. Using the operator form, we have [1]

$$\Phi^{(r+1)} = \frac{\mathcal{A}^{(r+1)^{-1}}(\Phi^{(r)})}{k_{\text{eff}}^{(r)}}\mathcal{P}\Phi^{(r)}, \quad k_{\text{eff}}^{(r+1)} = k_{\text{eff}}^{(r)}\frac{\langle\Phi^{(r+1)},\mathcal{P}\Phi^{(r+1)}\rangle}{\langle\Phi^{(r+1)},\mathcal{P}\Phi^{(r)}\rangle} \tag{6}$$

which is solved iteratively by power iterations. At every iteration, the removal matrix $\mathcal{A}$ contains the corrections given by the new estimate of the current by the integral expression. Hence, it can be written as $\mathcal{A}_0$ reproducing standard diffusion plus $\delta\mathcal{A}(\Phi)$ that contains the transport corrections. As mentioned, RM iterations must take into account the ordinary iteration schemes, like outers-inners. They can be implemented outside or inside the outers for instance, or even merge into a single iteration level. The inside-outer scheme is here employed.

*2.2. The drift current implementation*

The computational scheme of this algorithm is based on the introduction of the transport-corrected currents in the numerical scheme using drift terms

$$\delta J_{g,i+1/2} = J_{g,i+1/2} - J^D_{g,i+1/2} = -\delta D_{g,i+1/2}\frac{\Phi_{g,i+1}+\Phi_{g,i}}{(\Delta x_{i+1}+\Delta x_i)/2}. \tag{7}$$

In this case, the current acquires a contribution proportional to the flux, which is physically different from being proportional to the gradient of the flux, as in the previous implementation. The sum of the cell fluxes is proportional to the average local flux, representing a drift-advection term for neutron leakage. This option avoids indeterminate division by zero and it was originally introduced by the coarse mesh finite difference scheme (CMFD) in nodal diffusion codes [13]. For this implementation, we also need to specify the form for the current correction at the boundary, which has been defined as $\delta J_{g,-1/2} = -\delta D_{g,-1/2}\Phi_{g,0}$, without any division for the spatial width since no physical meaning is attributed to the correction itself [4]. The solving system is still expressed by

---

[1] $\mathcal{A}$ is a three diagonal banded matrix whose entries are the diffusion coefficient, plus removal terms for nuclear events. $\mathcal{P}$ is the neutron generator operator.

Eq. (6). A flow chart of the RM implementations is presented in Figure 1. The iterative algorithm, representative of both the implementations, is shown along with the Anderson acceleration performed through the DAAREM algorithm, which is crucial in finding the fixed point solution throughout the non-linear RM iterations [12].

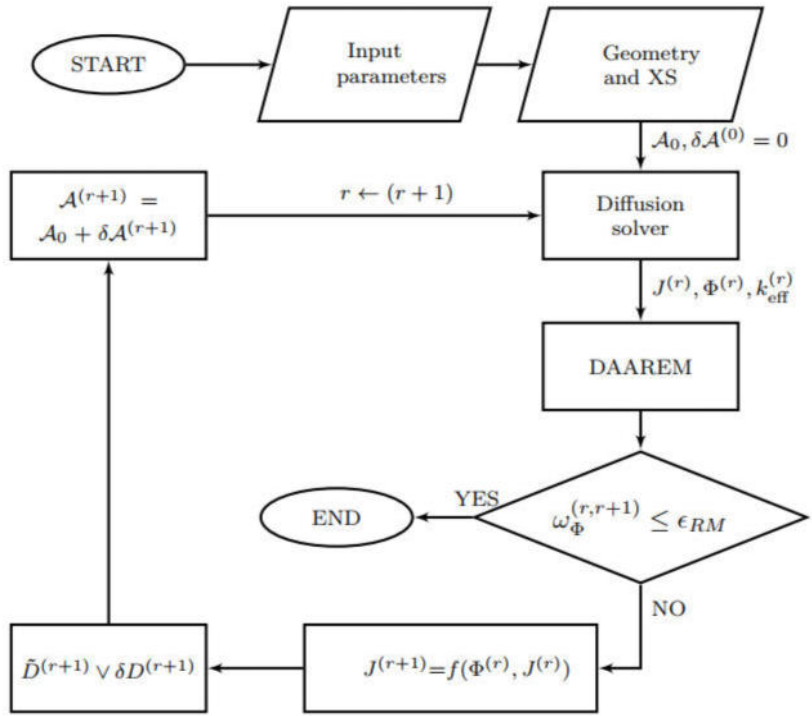

Figure 1: Flow chart of the RM.

The convergence criterion is based on the residuals of the relative flux differences between two successive iterations below a set threshold. Specifically, $\omega_\Phi^{(r,r+1)} = \max |(\Phi^{(r+1)} - \Phi^{(r)})/\Phi^{(r+1)}| \leq \epsilon_{RM}$. The computational drift scheme is also available through the implementation of the partial currents correction [5]. The correction terms in this scheme are two degrees of freedom (per interface per energy group) instead of one as done in the original drift implementation although no actual calculation benefit has been reported [5].

2.3. Boundary conditions

A generalized form of the boundary condition of (homogeneous) Robin type based on the use of the extrapolated distance $d_{ext,g}$ follows as $-d_x\Phi_g = \Phi_g/d_{ext,g}$. After the multiplication of the diffusion coefficient at both sides, this yields at the left boundary [12]

$$J_{g,-1/2}^D \simeq -\frac{D_{g,0}\Phi_{g,0}}{(\Delta_0/2 + d_{ext,g})} \qquad (8)$$

where $\Phi_{g,0}$ and $D_{g,0}$ are respectively the averaged flux and diffusion coefficient at the left border cell ($i = 0$), $\Delta_0/2$ is the half-width cell and $d_{ext,g} = \zeta_g D_{g,0}$ the extrapolated distance in the case of vacuum; usually, $\zeta_g = 2.13$ is the recommended value in homogeneous slab geometry [10]. Different boundary conditions can also be reproduced by this equation; for instance, reflection can be simulated by $\zeta_g \to \infty$ while the zero-flux comes with $\zeta_g = 0$.

Focusing on Eq. (8), the first cell-averaged flux $\Phi_{g,0}$ has been used as the boundary flux. The boundary value is not available according to the finite difference numerical scheme and hence an approximation is introduced. Studies related to a better approximation of the boundary flux have been investigated; specifically, we focused on a quadratic fit of the flux in the proximity of the boundary. The implementation is reported in Appendix A. This was a tentative to reduce the flux errors due to the approximation by discretization of the finite difference scheme. However, no major improvement concerning previous results is found when using higher-order numerical approximations of the boundary equation.

### 2.4. Inherent limitation of the $P_1$ approximation at vacuum boundary

When dealing with vacuum interfaces, it is worth recalling that an intrinsic limitation of diffusion theory in approximating the vacuum boundary exists. This is crucial for understanding the actual limitations of diffusion in modelling the vacuum boundary. Let us consider a finite 1D slab system of width $a$ with vacuum at the right boundaries. The actual transport boundary condition implies that the angular flux, $\psi(a,\mu)$, for all entering directions, is zero. In elementary diffusion theory, it is intuitive that we cannot satisfy this condition rigorously since we only deal with the first two flux moments. In diffusion, we accordingly require that the inwardly directed partial current $J_-(a)$ is vanishing for instance on the right surface. Specifically, for $\mu < 0$, we have the following

$$J_-(a) = \int_{-1}^{0} d\mu \mu \psi(a,\mu) = 0. \tag{9}$$

Diffusion theory can only yield an approximate description of the angular flux and therefore of the corresponding boundary condition. Recalling the $P_n$ theory, we can express the angular flux at the boundary as [3]

$$\psi(a,\mu) = \sum_{l=0}^{n} \left(\frac{2l+1}{2}\right) \Phi_l(a) P_l(\mu) \tag{10}$$

which for $n = 1$ leads to the diffusion approximation; hence, we can write

$$\psi(a,\mu) \simeq \frac{1}{2}\Phi(a) + \frac{3}{2}J(a)\mu, \tag{11}$$

where we identify the first two moments of the flux expansion as the scalar flux and the neutron current, respectively. Substituting the previous to the zero incoming current condition of Eq. (9) and performing the integration on $\mu$ leads to the following

$$\Phi(a) = 2J(a), \tag{12}$$

which is the well-known Marshak condition [3]. If substituted to Eq.(11), the following approximated expression of the angular flux at the boundary is expressed as

$$\psi(a,\mu) \simeq \frac{1}{2}\Phi(a)\left(1 + \frac{3}{2}\mu\right). \tag{13}$$

This relation gives negative values of $\psi(a,\mu)$ for $-1 \leq \mu < -2/3$, which is of course unphysical. This proves that we cannot reproduce the transport boundary condition $\psi(a,\mu) = 0$ with the $P_1$ approximation. A popular mathematical expedient is to introduce the extrapolated distance, as done in Eq. (8). It must be stressed that with this approximation, the flux does not vanish outside the boundary and that diffusion gives a poor representation of the true flux near the boundary. With the RM, a more accurate flux distribution to transport at the boundary can be obtained through the non-linear iterations but the intrinsic limitation of the approximated boundary condition persists.

## 3. Equivalence and customized diffusion

The RM allows having new estimates of the diffusion coefficient using Fick's law with a more accurate estimation of the currents by a transport operator. In practice, the best diffusion coefficient would be the one satisfying exactly Fick's law with the same flux and current provided by transport. An optimal diffusion coefficient can be defined using Fick's law with the flux distribution obtained from a transport solver. Moreover, an optimal extrapolated distance can also be redefined with this distribution to correctly reproduce the vacuum boundary condition. These new quantities are only meant to verify equivalence between diffusion and transport in terms of flux distribution calculations since usually the solution is neither known *a priori* nor available. The goal is to overcome the limitations of diffusion theory and prove an equivalence between the two physical models. Indeed, this is among the goals of the RM.

### 3.1. Optimal diffusion coefficients

Considering for simplicity a multi-group homogeneous slab with isotropic scattering only, the optimal diffusion coefficient $D_g^{opt}$, according to Fick's law, can be defined as

$$D_g^{opt}(x) = -\frac{J_g(x)}{d_x[\Phi_g(x)]} = -\frac{\int_0^x dx' E_2[\tau_g(x,x')]q_{0,g}(x') - \int_x^a dx' E_2[\tau_g(x',x)]q_{0,g}(x')}{\sigma_g(\int_0^x dx' E_0[\tau_g(x,x')]q_{0,g}(x') - \int_x^a dx' E_0[\tau_g(x',x)]q_{0,g}(x'))} \quad (14)$$

where both the neutron current $J_g(x)$ and flux derivative $d_x[\Phi_g(x)]$ are expressed by transport operators with well-known integrand quantities. The first-order flux derivative in Eq. (14) is computed using Leibniz integral rule and the transport kernel is written by employing the exponential integral function with $\tau_g(x',x)$ the optical path length and $\sigma_g$ the total cross section [1]. We also notice that the denominator of Eq. (14) involves integrals with singular kernels if $x = x'$ since $E_0[\tau_g(x,x)] = e^{-\tau_g(x,x)}/\tau_g(x,x) \to \infty$ [1]. However, it can still be proved that those integrals are finite, as demonstrated in Appendix C. Although the expressions for the current and the flux derivative are obtained for the continuum, here we use a numerical solution for the scalar flux when calculating the neutron source in the above integrals. The scalar flux is obtained by resolving the 1D multi-group problems with the CPM. The full slab is considered because of the intrinsic limitation of the CPM in treating reflection at the centre of the slab [12]. Using the standard notation for the spatial discretization of the numerical solution, Eq. (14) can be discretized as follows

$$D_{g,i+1/2}^{opt} = -\frac{\sum_{j=0}^{I-1} q_{0,g,j} I_{g,i,j,2} - \sum_{j=i+1}^{I-1} q_{0,g,j} I_{g,i,j,2}}{\sum_{j=0}^{I-1} \sigma_{g,j} q_{0,g,j} I_{g,i,j,0} - \sum_{j=i+1}^{I-1} \sigma_{g,j} q_{0,g,j} I_{g,i,j,0}} \quad (15)$$

where

$$I_{g,i,j,n} = \int_{x_{j-1/2}}^{x_{j+1/2}} dx' E_n[\tau_g(x_{i+1/2}, x')] \text{ for } n = 0, 2. \quad (16)$$

The previous integrals in the cells are computed by quadrature formulae (e.g., the midpoint rule). The extension to the heterogeneous media is straightforward for the current, using the expressions of Eq. (2). For the first-order flux derivative instead, although an analytical expression is possible but rather awkward, we have approximated it by central finite differences with a very fine spatial discretization. For test cases with no steep flux gradients at material interfaces, this is an acceptable approximation.

Supposing again that the solution from transport is known, we can also define an optimal extrapolated coefficient, called $\zeta_g^{opt}$, using its definition

$$\zeta_g^{opt} = -\frac{\Phi_{g,-1/2}}{J_{g,-1/2}}, \quad (17)$$

where the index $i = -1/2$ marks the quantities at the left boundary. A similar definition holds for the right one. We also note that the determination of the extrapolated distance is case-dependent because of the scalar flux. Moreover, for small problems (few mean free paths or less), this distance is expected to have a stronger effect not only in the proximity of the boundary. These new optimal quantities are tested in Section 5.1 as input data with a standard diffusion solver without requiring RM iterations.

*3.2. Analytical diffusion coefficients*

Alternative formulations of the diffusion coefficient and extrapolated distance based on the derivation of the exact analytical solution of the transport solution obtained by Case's method can be introduced. This method leads to exact expressions for the neutron distribution and criticality conditions. These expressions depend on expansion coefficients which are shown to satisfy a Fredholm integral equation. However, here we only focus on the results of diffusion theory with the exact Milne-problem extrapolated distance, which correspond to the zeroth-order approximation of the Neumann series solution to the Fredholm equation. In fact, no diffusion coefficient can be derived for higher-order approximations. The complete derivation, including also transport corrections, was accurately derived by George J. Mitsis [7]. The goal of introducing alternative definitions is to understand if different initial diffusion constants, obtained by physical assumptions or analytical methods, can affect the solution despite the intrinsic iterative procedure of the RM. The derivation of the diffusion coefficient and extrapolated distance, indicated as $D_{\nu_0}$ and $\zeta_0$, is provided in Appendix C.

## 4. Status and analysis of the RM performance

The RM has been recently tested with a complete test suite of 1G and 2G homogeneous and heterogeneous slab problems with isotropic and linearly-anisotropic scattering [5]. These cases belong to a test set of analytical benchmarks for code verification [9]. Reference flux distributions have been obtained using both reference solutions from Sood, a CPM solver for isotropic problems and $S_n$ solver for anisotropic ones. For the sake of brevity, RM performances compared to diffusion are summarized in Table 1, with test cases labelled according to the energy group, media property and scattering source. Details on results for each of the criticality problems can be found in [5]. Using the RM, the maximal deviation in criticality is less than ten pcm, and the maximal deviation in the spatial distribution of the flux is less than 2% and located mostly at the boundary with vacuum. Results are representative of both the RM implementations. A convergence criterion is applied in the iterative solving scheme of the RM, as highlighted in Figure 1. Indeed, previous results were obtained with a value of $\epsilon_{RM} = 10^{-6}$. However, the effects of the convergence criterion have not been investigated in previous research. Lowering this threshold will eventually affect the convergence rate and require in theory more RM iterations but no effects on the flux distribution can be predicted. Results related to this investigation are shown in Section 5.2.

**Table 1**
Summary of RM performances (with standard $D_0, \zeta_0$): results for Sood test cases in slab geometry.

| Test case | ($\approx$) max$|(1-k)|$ (pcm) | ($\approx$) max$|(\Delta\Phi^{(RM)})|$ % |
|---|---|---|
| $\epsilon_{RM} = 10^{-6}, \Delta = 0.01$ mfp | **Diffusion / RM** | **Diffusion / RM** |
| *Isotropic cases* | | |
| 1G homogeneous | **12,000 / 4** | **30 / 1.2 (v.b.)** |
| 2G homogeneous | **14,000 / 7** | **80 / 1.1 (v.b.)** |
| 1G heterogeneous | **15,000 / 5** | **16 / 0.8 (m.i.)** |
| 2G heterogeneous | **300 / 3** | **25 / 1.5 (v.b)** |
| *Linearly-anisotropic cases* | | |
| 1G homogeneous | **10,000 / 2** | **30 / 1.4 (v.b.)** |
| 2G homogeneous | **18,000 / 3** | **75 / 1.3 (v.b.)** |

v.b. = vacuum boundary, m.i. = material interface.

## 5. Results

We here present the results of the investigations reported in Section 3 and Section 4. Results are shown for homogeneous and heterogeneous problems from Sood's analytical benchmarks test set [9]. Problem specifications are available in Appendix D.

*5.1. Transport-diffusion equivalence check*

The numerical solution of the diffusion equation is presented for homogeneous and heterogeneous problems[2] using different diffusion coefficients as inputs, as shown in Figure 3. A schematic representation of the slab configurations with reference flux positions is given in Figure 2. Standard diffusion coefficients are denoted with $D_0 = 1/3\sigma_0$ and $\zeta_0 = 2.13$. The analytical diffusion coefficients for 1G homogeneous cases only are referred to as $D_{\nu_0}$ and $\zeta_{\nu_0}$. Finally, the optimal coefficients are labelled as $D^{opt}$ and $\zeta^{opt}$.

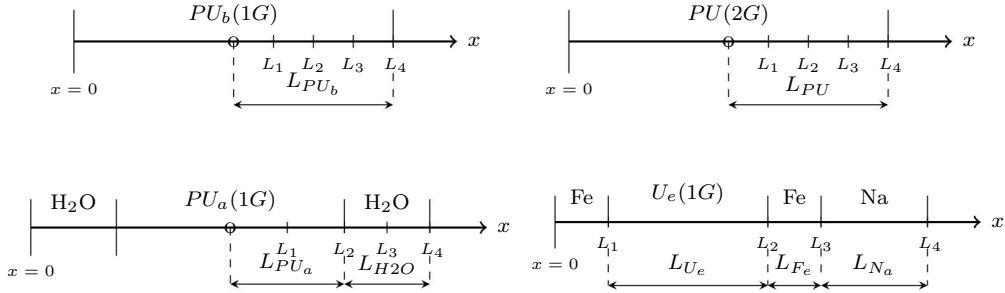

Figure 2: Geometrical specifications for 1G-2G homogeneous and 1G heterogeneous test cases.

A very fine spatial discretization, $\Delta = 0.001$ mfp, has been used in all calculations. The solution by standard diffusion in these particular test cases are very different from the reference transport solution provided by the CPM. All flux deviations are computed as $\Delta\Phi\% = (1 - \Phi/\Phi_{CPM}) * 100$, whether the flux is computed by a standard diffusion calculation or the RM.

---

[2]The heterogeneous test case Ue-Fe-Na-1-0-SL is only employed for evaluating the second part of the results. No equivalence is provided for this case due to very steep flux gradients in the computation of $D^{opt}$, invalidating the procedure.

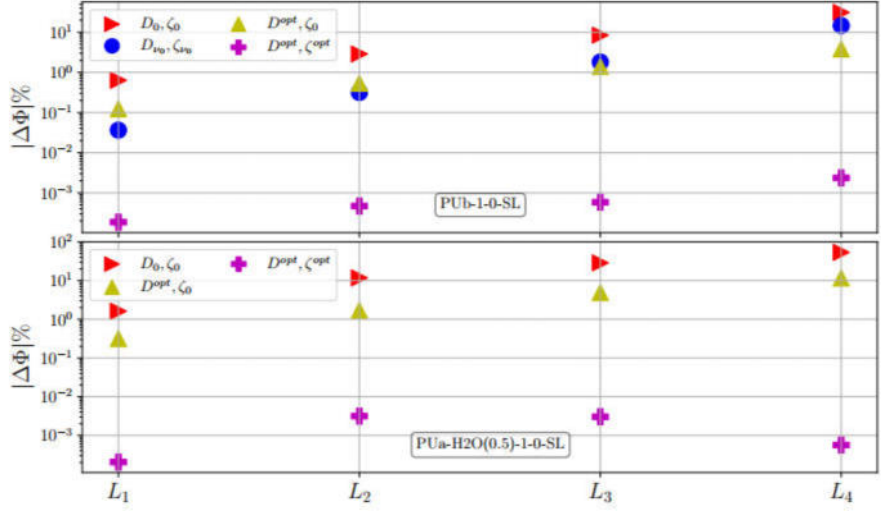

(a) 1G test cases.

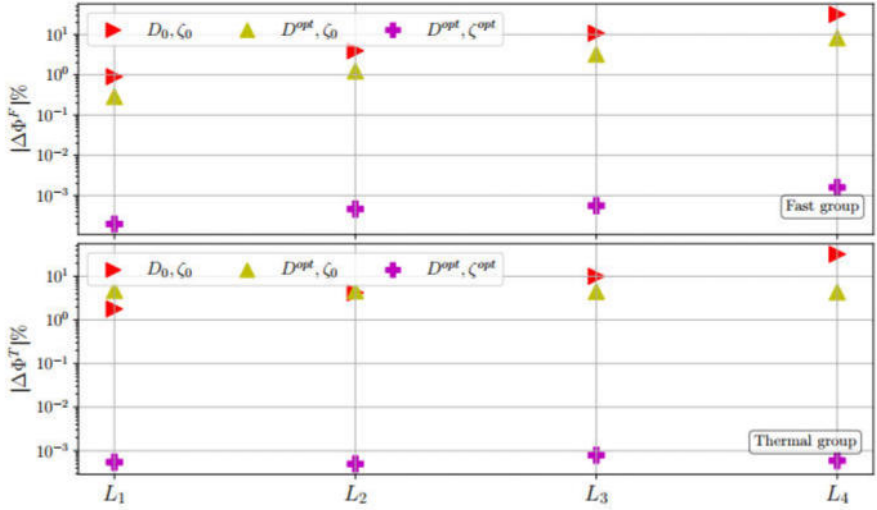

(b) 2G test case: PU-2-0-SL.

Figure 3: Flux deviation (%) using diffusion with customized diffusion constants.

The use of standard diffusion coefficients $D_0$, $\zeta_0$ results in flux deviations of around tens of percent as we approach the vacuum boundaries and other material interfaces (e.g, reflector). For the 1G homogeneous problem, the use of the analytical diffusion constants results in a better agreement to transport but considerably large flux deviations are still present in the proximity of the vacuum boundary. A similar trend is obtained when the optimal diffusion coefficient $D^{opt}$ is used along with $\zeta_0$. This behaviour is due to the improper extrapolated coefficient used to represent the vacuum boundary condition. The solution achieves an accuracy of at least four decimal places with the CPM solution at all reference positions when $\zeta^{opt}$ is employed. These results prove that by employing optimal coefficients an equivalence is set between the transport and diffusion model. In Figure 4 and 5, the flux deviations at the boundary with vacuum are shown using $D^{opt}$ with a range of extrapolated coefficients. The error is minimized when $\zeta^{opt}$ is employed, which means that the vacuum has been correctly modelled by the extrapolated distance $\zeta^{opt} D^{opt}$. The optimal extrapolated distance is also unique in minimizing the flux error. Small deviations from this value cause a significant increase in the flux deviation. Modelling correctly the vacuum boundary is essential in obtaining an equivalence between the two at all reference positions with a standard diffusion calculation.

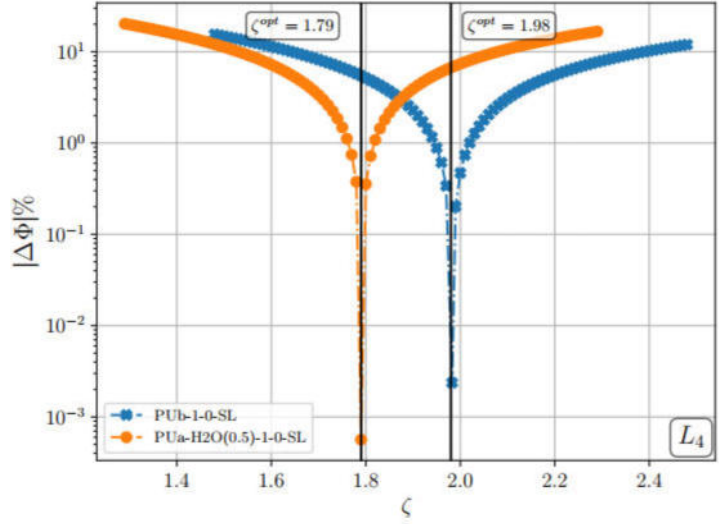

Figure 4: Flux deviation (%) trends with extrapolated coefficient using $D^{opt}$ for 1G test cases.

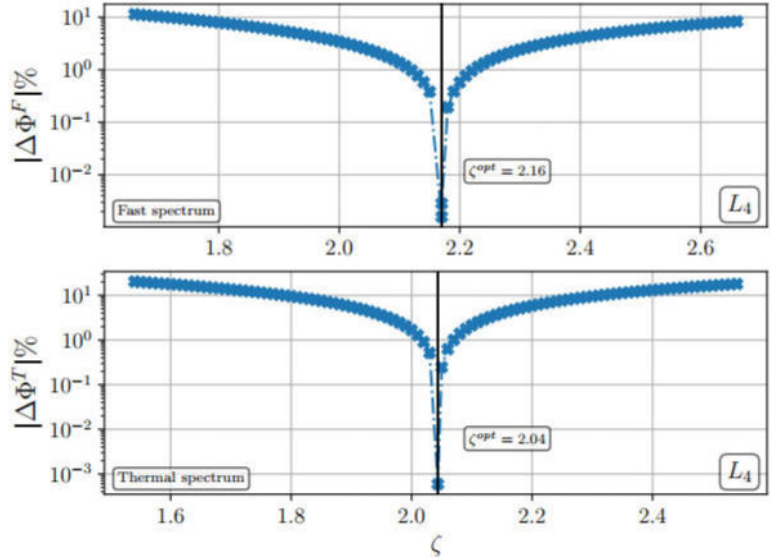

Figure 5: Flux deviation (%) trends with extrapolated coefficient using $D^{opt}$ for 2G test case.

### 5.2. On the convergence criterion in RM iterations

Both RM implementations are employed and compared using standard diffusion coefficients as input data with decreasing convergence criterion for all the test cases introduced in section 5.1. The CPM is used again as the reference solution. Flux percentage deviations at the vacuum boundary and material interfaces are presented in Figure 6, 7 and 8. The spatial mesh of 0.001 mfp has been used. In addition, analytical coefficients derived in section 3.2 are also employed for the 1G homogeneous problem. Tightening the convergence criterion does reduce monotonically the flux deviations, which are lowered down to $\sim 0.1\%$ and $\sim 0.01\%$ for the $RM^{net}$ and $RM^D$ implementation with a minimum tolerance of $10^{-10}$, respectively. The 2G problem's flux deviation reached even lower values. Hence, an improvement of at least one and two orders of magnitude in the flux deviation at the vacuum boundary is obtained with respect to previous results [5]. The $RM^D$ implementation has generally shown a more accurate flux distribution evaluation at the vacuum boundary. At material interfaces but the boundaries, no difference between the two implementations has been noticed, as highlighted in Figure 8.

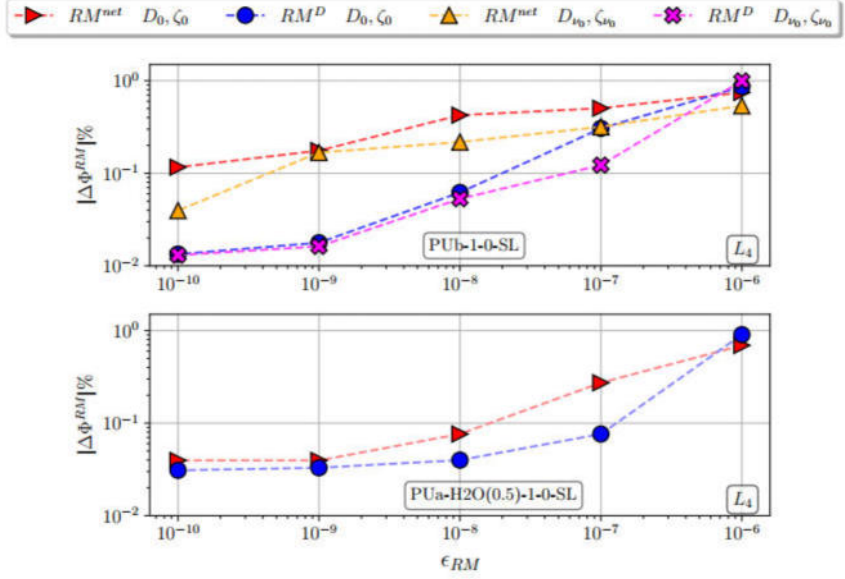

Figure 6: Flux % deviations trends using RM with decreasing $\epsilon_{RM}$ (1G test cases).

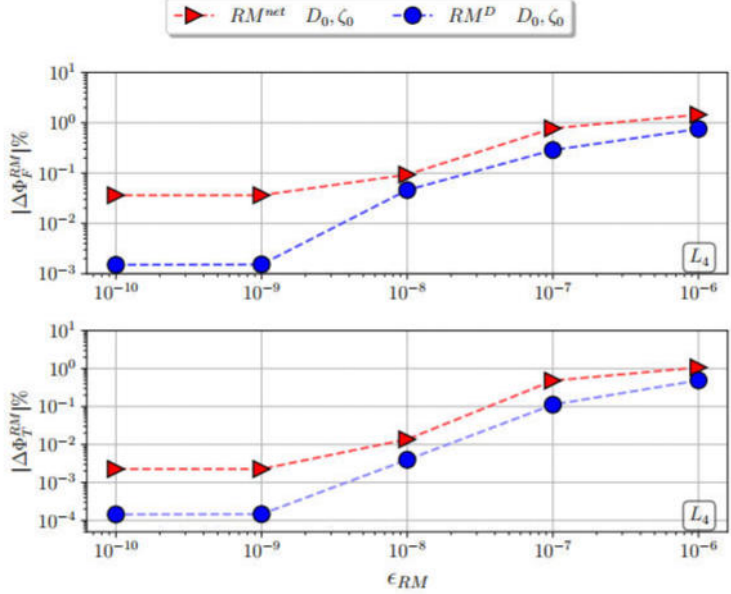

Figure 7: Flux % deviations trends using RM with decreasing $\epsilon_{RM}$ (2G test case).

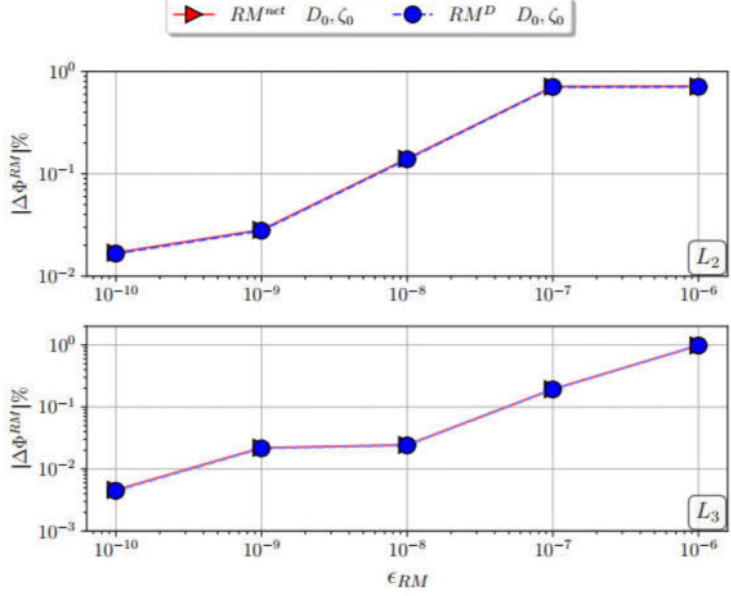

Figure 8: Flux % deviations trends at material interfaces using RM with decreasing $\epsilon_{RM}$ (Ue-Fe-Na-1-0-SL).

In addition, with this implementation, no differences in the flux deviations are found by using different input diffusion coefficients (e.g, $D_0, \zeta_0$; $D_{\nu_0}, \zeta_{\nu_0}$) at the beginning of the iterative scheme after a value of $\epsilon_{RM} = 10^{-8}$, as reported in Figure 6. Indeed, the right diffusion coefficients are in practice not needed at the beginning of the diffusion calculation if tighter tolerances are used. This confirms and completes the intuition by Tomatis and Dall'Osso on the advantage of such an iterative scheme in the first numerical application of RM [11]. The number of iterations to achieve the previous flux deviations is reported in Table 2. The increase in the number of RM iterations is relatively acceptable compared to the gain in flux accuracy. Once again, the use of the DAAREM algorithm[3] as an acceleration scheme applied to RM iterations is crucial in obtaining fast convergence, especially with the use of a small convergence criterion.

**Table 2**

Number of RM iterations with convergence criteria for 1G-2G test cases.

| Test case | Diff. coeffs (RM*) | $\epsilon_{RM}$ | | | | |
|---|---|---|---|---|---|---|
| | | $1e-6$ | $1e-7$ | $1e-8$ | $1e-9$ | $1e-10$ |
| PUb-1-0-SL | $D_0, \zeta_0$ (RM$^{net}$) | 123 | 220 | 238 | 427 | 569 |
| | $D_{\nu_0}, \zeta_{\nu_0}$ (RM$^{net}$) | 102 | 223 | 230 | 323 | 682 |
| | $D_0, \zeta_0$ (RM$^D$) | 155 | 308 | 497 | 572 | 746 |
| | $D_{\nu_0}, \zeta_{\nu_0}$ (RM$^D$) | 81 | 320 | 432 | 557 | 658 |
| PUa-H2O(0.5)-1-0-SL | $D_0, \zeta_0$ (RM$^{net}$) | 79 | 459 | 545 | 572 | 1120 |
| | $D_0, \zeta_0$ (RM$^D$) | 138 | 458 | 545 | 572 | 1141 |
| PU-2-0-SL | $D_0, \zeta_0$ (RM$^{net}$) | 184 | 722 | 723 | 1195 | 1296 |
| | $D_0, \zeta_0$ (RM$^D$) | 242 | 334 | 868 | 1238 | 1442 |
| Ue-Fe-Na-1-0-SL | $D_0, \zeta_0$ (RM$^{net}$) | 181 | 188 | 488 | 995 | 1078 |
| | $D_0, \zeta_0$ (RM$^D$) | 177 | 380 | 572 | 626 | 1046 |

---

[3]An improved version of the DAAREM has been used to generate these results. Remember to reference.

*5.3. Diffusion coefficient trends in the proximity of material interfaces*

The optimal diffusion quantities defined in section 3.1 were computed using the flux distribution obtained from the CPM. However, the flux distribution is usually not known and this check was intentionally set to verify the equivalence between diffusion and transport for very simple test cases. Nevertheless, those coefficients can be considered the reference ones by which comparisons can be made with the corrected diffusion coefficients from RM calculations. In section 5.1, we also saw how the extrapolated distance can affect the flux distribution close to the vacuum boundary. In the RM, the extrapolated coefficient $\zeta$ is kept constant throughout the iterations and set equal to the suggested value of $\zeta_0 = 2.13$ [10]. However, a new flux distribution is available at every iteration and the extrapolated distance can be redefined accordingly to its definition using the latest flux distribution. At the end of every RM iteration, $d_{ext}$ can be redefined using the redefinition of the diffusion coefficient implementation since a corrected diffusion coefficient is needed.

To get some additional insight regarding this possibility, it is interesting to visualize the diffusion coefficient trends at the end of Ronen iterations in the proximity of vacuum. Figure 9 shows the trends of $D_0$ with and without RM in comparisons with $D^{opt}$ for a 1G homogeneous case. Diffusion coefficient trends at the material interface (reflector) are also shown for a 1G heterogeneous test case in Figure 10. Similar trends are obtained for the other test cases without providing further information. When the RM is employed, trends are also shown with respect to decreasing values of the convergence criterion. The vacuum boundary condition has been modelled with two extrapolated coefficients: $\zeta_0$ and $\zeta^{opt}$. The way we model the vacuum boundary does affect the diffusion coefficient at this interface, leading to negative and positive values of the diffusion coefficient for $\zeta_0$ and $\zeta^{opt}$, respectively. Using $\zeta_0$ results in having a negative extrapolated distance $d_{ext}$, which is unphysical and prevents the redefinition of extrapolated distance itself. However, having negative corrections at the vacuum boundary does not prevent obtaining very accurate flux distributions. Using $\zeta^{opt}$ instead leads to positive diffusion coefficients everywhere but its value is usually unknown. The error trends at the vacuum boundary using the three extrapolated coefficients for the $RM^D$ implementation are shown in Table 3. The last table clearly shows that the solution does converge to the same distribution for tight convergence criteria regardless of the way we model the vacuum boundary. Hence, we conclude that for very tight tolerances, the value of the diffusion coefficient at the boundary is practically irrelevant, whether it is physical or not. This also shows that the reference solution can be achieved by RM with multiple pairs of $D$ and $\zeta$ when tight convergence tolerances are set. As mentioned above, redefining the extrapolated coefficient although correct in theory, brings in some numerical problems related to negative extrapolated distances and is practically irrelevant for tight convergence criteria.

**Table 3**
Flux deviation (%) at vacuum boundary with extrapolated coefficients using RM for PUb-1-0-SL test case.

| $\epsilon_{RM}$ | $\Delta\Phi_{L_4}\%$ ($D_0, \zeta_0 = 2.13$) | $\Delta\Phi_{L_4}\%$ ($D_0, \zeta_{\nu_0} = 1.57$) | $\Delta\Phi_{L_4}\%$ ($D_0, \zeta^{opt} = 1.98$) |
|---|---|---|---|
| 1.e-6 | 0.769833 | 0.490860 | 0.519612 |
| 1.e-7 | 0.212684 | 0.054975 | 0.040269 |
| 1.e-8 | 0.07456 | 0.064199 | 0.058997 |
| 1.e-9 | 0.03247 | 0.064199 | 0.025689 |
| 1.e-10 | **0.02247** | **0.019988** | **0.023171** |

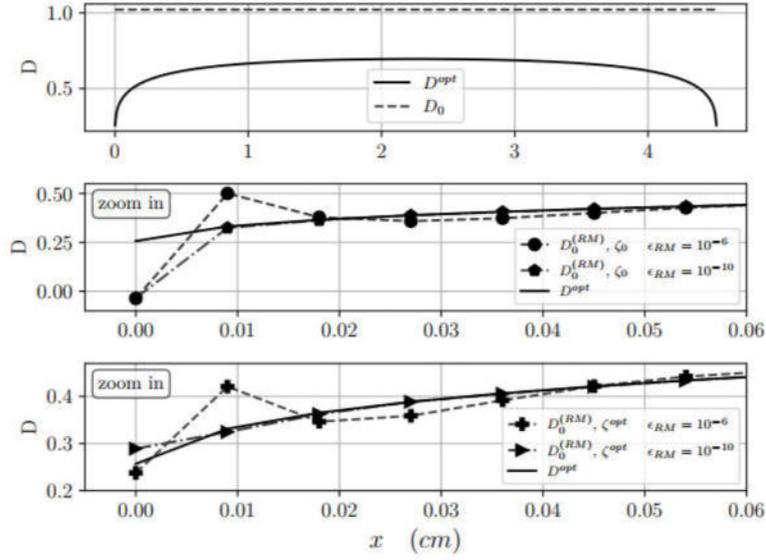

Figure 9: Diffusion coefficient (cm) trends for PUb-1-0-SL test case.

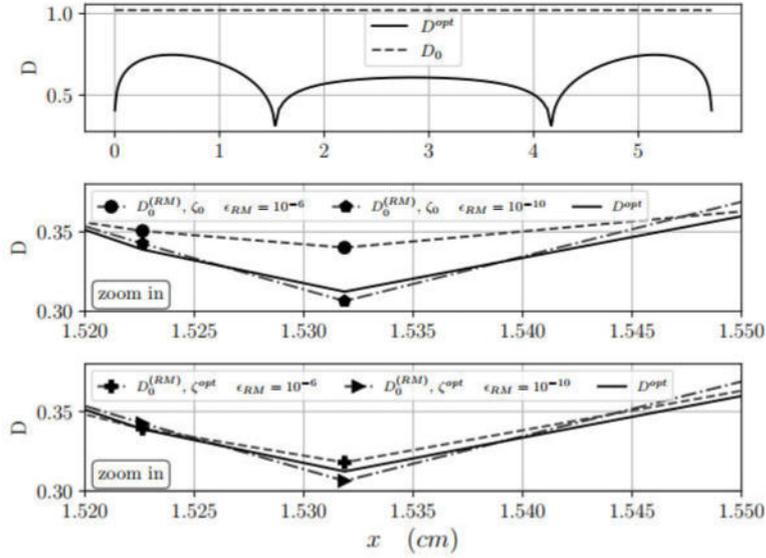

Figure 10: Diffusion coefficient (cm) trends for PUa-H2O(0.5)-1-0-SL test case.

## 6. Conclusion

The RM has been recently tested for a wide range of problems in 1D plain geometry, including anisotropic scattering test cases [5]. Some discrepancies on the scalar flux at the material interface (especially with vacuum) have been observed, as previously reported for simple cases in [5, 12]. The latest implementation also includes the use of the Anderson acceleration through the DAAREM algorithm, providing fast calculations. However, these results need further numerical and physical investigations.

In this article, we provide investigations carried out to explain these flux deviations. The use of higher-order numerical approximations of the discretized equation at the bound-ary does not provide any considerable improvement. Hence, physical and numerical investigations are set. After recalling the intrinsic limitation of diffusion theory in modelling the vacuum boundary, optimal diffusion coefficients are defined and tested with standard diffusion for 1G-2G test cases. Along this, an analytical expression for the diffusion coefficient and extrapolated distance using Case's method are obtained and tested for a 1G homogeneous problem. Performances using standard diffusion with these parameters are presented. An equivalence between transport and diffusion is verified with the use of optimal coefficients. The RM has shown high sensitivity of the iterative method to the convergence tolerance. Tightening the algorithm tolerance leads to very accurate flux distribution at every material interface with an accuracy of at least three decimals and beyond respect to the reference solution, regardless of the modelling of the vacuum with the extrapolated distance. This is a significant improvement compared to previous results, although an increasing number of RM iterations is reported.

At very tight algorithm tolerances, the difference in the diffusion coefficients at the boundary seems to not affect the flux distribution. In conclusion, the causes of the discrepancy in the flux distribution at material interfaces have been explained. This has led to a further evaluation of the RM performances. After providing these essential studies in the methodology and performance, the upgrade of the RM to a true coarser meshes implementation will be the object of future research which will determine its advancement to practical applications.

**Acknowledgments**

The authors are thankful to Framatome for supporting financially this study as part of a collaboration agreement regarding this research (Grant number: FRA-21-002-RM).

**Appendix A. Higher-order approximation for the boundary equation**

To compute a more accurate current at the vacuum boundary, here we look for a better estimation of the boundary flux indicated as $\Phi_b$, which is currently assumed to be the cell-averaged, as stated in Eq. (8). We suppose that the flux can be written as a quadratic polynomial function $\Phi(x) = a_0 + a_1 x + a_2 x^2$ with $x$ the distance from the left boundary and the unknown $a_0, a_1, a_2$ to be determined. The following conditions are set

$$\Phi(0) = \Phi_b = a_0$$
$$\Phi'(0) = a_1 = \frac{\Phi_b}{\zeta D} = \frac{a_0}{\zeta D} \quad (A.1)$$
$$\int_{x_i}^{x_{i+1}} \Phi(x)dx = \Phi_i \Delta_{i,i+1} = \Phi_b \Delta_{i,i+1} + \frac{\Phi_b}{\zeta D} \frac{x^2}{2}\Big|_{x_i}^{x_{i+1}} + a_2 \frac{x^3}{3}\Big|_{x_i}^{x_{i+1}}$$

with $\Delta_{i,i+1} = x_{i+1} - x_i$ the spatial width. The second condition refers to the zero incoming current condition whereas in the third we preserve the average values in the first two cells near the boundary by applying the mean value theorem. Referring to the first two cells coordinates as $[x_0, x_1]$ and $[x_1, x_2]$, we get after rearrangement

$$\Phi_0 = \Phi_b + \frac{\Phi_b}{\zeta D}\bar{x}_{10} + \frac{a_2}{3}(4\bar{x}_{10}^2 - x_0 x_1) \quad (A.2)$$

and

$$\Phi_1 = \Phi_b + \frac{\Phi_b}{\zeta D}\bar{x}_{21} + \frac{a_2}{3}(4\bar{x}_{21}^2 - x_1 x_2), \quad (A.3)$$

with $\bar{x}_{10} = (x_1 + x_0)/2$ and $\bar{x}_{21} = (x_2 + x_1)/2$. Subtracting the previous two and isolating $a_2$ we find the last coefficient

$$a_2 = 3\frac{\left[(\Phi_1 - \Phi_0) - \frac{\Phi_b}{\zeta D}(\bar{x}_{21} - \bar{x}_{10})\right]}{4(\bar{x}_{21}^2 - \bar{x}_{10}^2) + x_1 x_0 - x_2 x_1}. \quad (A.4)$$

Hence, our polynomial flux approximation can be written as

$$\Phi(x) = \Phi_b + \frac{\Phi_b}{\zeta D}x + 3\frac{\left[(\Phi_1 - \Phi_0) - \frac{\Phi_b}{\zeta D}(\bar{x}_{21} - \bar{x}_{10})\right]}{4(\bar{x}_{21}^2 - \bar{x}_{10}^2) + x_1 x_0 - x_2 x_1}x^2 \quad (A.5)$$

which is a function of the unknown boundary flux $\Phi_b$ and average fluxes $\Phi_0, \Phi_1$. Isolating for instance $\Phi_b$ in Eq. (A.2) and substituting the expression of $a_2$, we get an expression for the boundary flux

$$\Phi_b = \frac{\Phi_0 - \frac{(\Phi_1 - \Phi_0)(4\bar{x}_{10}^2 - x_0 x_1)}{4(\bar{x}_{21}^2 - \bar{x}_{10}^2) + x_1 x_0 - x_2 x_1}}{c} \quad (A.6)$$

with $c$ defined by geometry and medium quantities only

$$c = 1 + \frac{\bar{x}_{10}}{\zeta D} - \frac{(\bar{x}_{21} - \bar{x}_{10})(4\bar{x}_{10}^2 - x_0 x_1)}{\zeta D\left[4(\bar{x}_{21}^2 - \bar{x}_{10}^2) + x_1 x_0 - x_2 x_1\right]}. \quad (A.7)$$

We are now able to express the polynomial fit of the flux by only unknown average fluxes $\Phi_0$ and $\Phi_1$, which can be added to the solving diffusion solver. Factorizing the solution by the average fluxes $\Phi_0$ and $\Phi_1$, we get

$$\Phi(x) = \left[\frac{1}{c}(1+q)\left(1+\frac{1}{\zeta D}x\right) - 3\left(1 + \frac{q(\bar{x}_{21} - \bar{x}_{10})}{c\zeta D} + \frac{(\bar{x}_{21} - \bar{x}_{10})}{c\zeta D}\right)qx^2(4\bar{x}_{10}^2 - x_0 x_1)\right]\Phi_0$$
$$+ \left[-\frac{q}{c}\left(1+\frac{1}{\zeta D}x\right) + 3\left(1 + \frac{q(\bar{x}_{21} - \bar{x}_{10})}{c\zeta D}\right)qx^2(4\bar{x}_{10}^2 - x_0 x_1)\right]\Phi_1, \quad (A.8)$$

with

$$q = \frac{4\bar{x}_{10}^2 - x_0 x_1}{(4(\bar{x}_{21}^2 - \bar{x}_{10}^2) + x_1 x_0 - x_2 x_1)}. \quad (A.9)$$

For the left border ($x = 0$), the expression simplifies to

$$\Phi_b = \Phi(0) = \left(\frac{1}{c} + \frac{4\bar{x}_{10}^2 - x_0 x_1}{c(4(\bar{x}_{21}^2 - \bar{x}_{10}^2) + x_1 x_0 - x_2 x_1)}\right)\Phi_0$$
$$- \left(\frac{4\bar{x}_{10}^2 - x_0 x_1}{c(4(\bar{x}_{21}^2 - \bar{x}_{10}^2) + x_1 x_0 - x_2 x_1)}\right)\Phi_1. \quad (A.10)$$

A simple change of variable $x = a - x$ is required for the right border. Results showed that very few improvements are detected with the higher-order approximation since very fine spatial meshes were already used in previous research [5, 12].

## Appendix B. Regularization of singular integrals

The reference diffusion coefficient has been introduced in section 3.1 using Fick's law with the flux derivative expressed as

$$d_x[\Phi_{tr}(x)] = \sigma_t\left(\int_0^x dx' E_0[\tau(x, x')]\Phi_{tr}(x') - \int_x^a dx' E_0[\tau(x', x)]\Phi_{tr}(x')\right). \quad (B.1)$$

Here, the kernel has been written using the exponential integral function [1]

$$E_n(\tau) = \int_0^1 d\mu\, e^{-\tau/\mu}\mu^{n-2} = \int_1^\infty du\, e^{-\tau u}u^{-n}, \; n \geq 0 \quad (B.2)$$

with $\tau(x', x)$, the optical path length, defined as $\tau(x', x) = Sgn(x - x')\int_{x'}^x \sigma_t(x'')dx''$. We do notice that this expression involves integrals with singular kernels when $x = x'$ since $E_0[\tau(x,x)] = e^{-\tau(x,x)}/\tau(x,x) \to \infty$. However, it can still be proved that those integrals are finite. Isolating the singularity in zero, we have

$$\sigma_t\left(\int_0^x dx' E_0[\tau(x,x')]\Phi_{tr}(x') - \int_x^a dx' E_0[\tau(x',x)]\Phi_{tr}(x')\right) =$$
$$\sigma_t \lim_{\varepsilon \to 0}\left(\int_0^{x-\varepsilon} dx' E_0[\tau(x,x')]\Phi_{tr}(x') + \int_{x-\varepsilon}^{x+\varepsilon} dx' E_0[\tau(x,x')]\Phi_{tr}(x') - \int_{x+\varepsilon}^a dx' E_0[\tau(x',x)]\Phi_{tr}(x')\right) \quad (B.3)$$

and using the derivative property of the exponential integral function [1]

$$\frac{dE_1(z)}{dz} = -E_0(z), \quad (B.4)$$

with $z = \tau(x', x)$, the second term of the RHS of Eq. (B.3) can be recast as

$$\sigma_t \lim_{\varepsilon \to 0}\int_{x-\varepsilon}^{x+\varepsilon} dx' E_0[\tau(x',x)]\Phi(x') \simeq \lim_{\varepsilon \to 0}\bar{\Phi}(x)\left(E_1[\tau(x+\varepsilon, x)] - E_1[\tau(x-\varepsilon, x)]\right). \quad (B.5)$$

In the previous, we introduced an average finite flux $\bar{\Phi}(x)$ in the domain of the singularity. For $\varepsilon \to 0$, the $E_1$ terms go to infinity with the same speed but with an opposite sign; hence, due to the odd singularity in zero, the limit itself is zero and the integral stays finite.

## Appendix C. The derivation of analytical diffusion coefficients

Considering a one-energy group and one-dimensional uniform multiplying media, the angular flux $\psi(x,\mu)$ satisfies the following transport equation

$$\mu \frac{\partial \psi(x,\mu)}{\partial x} + \psi(x,\mu) = \frac{c}{2}\int_{-1}^{1}\psi(x,\mu')d\mu', \tag{C.1}$$

where $c = (\sigma_s + \nu\frac{\sigma_f}{k_{\text{eff}}})/\sigma$ is defined as the mean number of secondary neutrons produced per neutron reaction. The distance $x$ is in units of the mean free path. According to Case's method, the general solution of Eq. (C.1) can be found by expanding $\psi(x,\mu)$ in an appropriate set of singular eigenfunctions, for instance,

$$\psi(x,\mu) = e^{-x/\nu}\phi_\nu(\mu) \tag{C.2}$$

where $\phi_\nu(\mu)$ satisfies the following equation

$$1 - (\frac{\mu}{\nu})\phi_\nu(\mu) = \frac{c}{2}\int_{-1}^{1}\phi_\nu(\mu')d\mu', \tag{C.3}$$

which can be seen as an eigenvalue integral equation where $\nu$ plays the role of the eigenvalue. Using the normalization condition $\int_{-1}^{1}\phi_\nu(\mu)d\mu = 1$, which is arbitrary because of the homogeneity of the equation, we obtain the following expression for the eigenfunction [2]

$$\phi_\nu(\mu) = \frac{c}{2}P\frac{\nu}{\nu-\mu} + \lambda(\nu)\delta(\mu-\nu) \tag{C.4}$$

where P denotes principal values of the integrals of $\phi_\nu(\mu)$ and the second term appears only for singularity points ($\nu = \mu$). In theory, $\nu$ can take any value inside the domain of $[-1,1]$ where $\mu$ is defined as well as outside this domain; for $\nu \notin [-1,1]$, only the first term of Eq. (C.3) is present with $P = 1$. In this last occurrence, the arbitrary normalization condition determines the value of the eigenvalue $\frac{1}{c\nu} = tanh^{-1}(1/\nu)$ which is a transcendental equation in $\nu$. This transcendental equation can be easily solved by a root-finding method (e.g., bisection); two purely imaginary roots, denoted as $\pm\nu_0$, are found for $c > 1$ as shown in Figure C.11. The corresponding eigenfunctions are defined as

$$\phi_{0\pm}(\mu) = \frac{c}{2}\frac{\nu_0}{\nu_0 \mp \mu}. \tag{C.5}$$

For $\nu$ belonging to $[-1,1]$, the normalization condition simply defines the form of $\lambda(\nu)$, leaving $\nu$ unrestricted with $\phi_\nu(\mu)$ given by Eq. (C.4). Recalling the separation of variables in Eq. (C.2), the general solution can be finally written as

$$\psi(x,\mu) = a_{0+}\phi_{0+}(\mu)e^{-x/\nu_0} + a_{0-}\phi_{0-}(\mu)e^{+x/\nu_0} + \int_{-1}^{1}A(\nu)\phi_\nu(\mu)e^{-x/\nu}d\nu, \tag{C.6}$$

where the first two terms at the RHS are entitled as the discrete eigenfunctions, while the third term is the continuum solution. For the latter, an integration over the continuum spectrum is necessary over the continuous variable $\nu$. The expansion coefficients $a_{0+}, a_{0-}$ and $A(\nu)$ are determined according to the applied boundary conditions. Orthogonality of the eigenfunctions is also used to obtain these coefficients. Considering a 1D slab configuration of half-width $b$ with the origin of the coordinate at the centre slab, expansion coefficients are obtained from the boundary and symmetry conditions. Considering vacuum at the boundaries and symmetric properties, we have $\psi(x,\mu) = \psi(-x,-\mu)$ and $\psi(-b, u > 0) = 0$ [7]. Substituting the general solution of Eq. (C.6) in the first of the previous relationship and using the symmetry properties of the eigenfunctions $\psi_{0\pm}(-x,-\mu) = \psi_{0\mp}(x,\mu)$, $\psi_\nu(-x,-\mu) = \psi_{-\nu}(x,\mu)$ we get

$$(a_{0+} - a_{0-})\psi_{0+}(x,\mu) + (a_{0-} - a_{0+})\psi_{0-}(x,\mu) + \int_{-1}^{1}[A(\nu) - A(-\nu)]\psi_\nu(x,\nu)d\nu = 0. \tag{C.7}$$

The previous one holds for both the entire spatial and angular domain, hence we conclude that the expansion coefficients must satisfy $a_{0+} = a_{0-}$ and $A(\nu) = A(-\nu)$. We anticipate the result of Mitsis which proved that the expansion coefficient $A(\nu)$ obeys the Fredholm integral equation which can be solved with an approximation procedure based on the Neumann series solution [7]. In addition, no constraint on $a_{0+}$ is found and hence it remains arbitrary. We note that neglecting the derivation of $A(\nu)$ is here justified by the fact that for the diffusion solution, no contribution from the continuum solution is needed, as it will be demonstrated. We proved that the angular flux can be written as

$$\psi(x,\mu) = a_{0+}[\psi_{0+}(x,\mu) + \psi_{0-}(x,\mu)] + \int_{-1}^{1} A(\nu)\psi_{\nu}(x,\mu)d\nu. \tag{C.8}$$

We can now obtain the neutron flux from its definition, $\Phi(x) = \int_{-1}^{1} \psi(x,\mu)d\mu$ which by applying the normalization condition gives

$$\Phi(x) = a_{0+}[e^{-x/\nu_0} + e^{+x/\nu_0}] + \int_{-1}^{1} A(\nu)e^{-x/\nu}d\mu. \tag{C.9}$$

Since the discrete eigenvalue is purely imaginary ($c > 1$) and the continuous coefficient $A(\nu)$ is an even function, the scalar flux can be written as

$$\Phi(x) = 2a_{0+}cos(x/|\nu_0|) + 2\int_{0}^{1} A(\nu)cosh(x/\nu)d\mu. \tag{C.10}$$

Analogously for the neutron current, following its definition and the normalization condition, we get

$$J(x) = 2a_{0+}(c-1)|\nu_0|sin(x/|\nu_0|) + 2(c-1)\int_{0}^{1} A(\nu)\nu sinh(x/\nu)d\mu. \tag{C.11}$$

We immediately notice that the first term of the flux is identified with the asymptotic solution from diffusion and that both the flux and the current have the expected symmetry property. The flux is symmetric with respect to the origin while the current vanishes at the origin. By looking carefully at Eqs. (C.10) and (C.11), not considering the continuum contribution, we can write the following relation between the two quantities

$$J_0(x) = -D_{\nu_0}\frac{d\Phi_0(x)}{dx} \tag{C.12}$$

where the index 0 has been added to recall that it was obtained from the discrete solution. This expression, which is exactly Fick's law, clearly states the proportionality between the current and the flux, with the diffusion coefficient defined as $D_{\nu_0} = |\nu_0|^2(c-1)$.

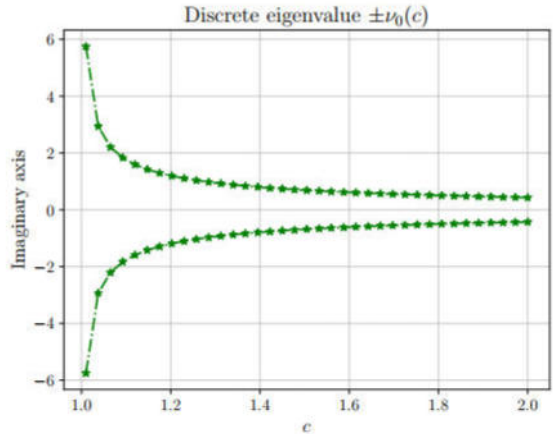

Figure C.11: Discrete eigenvalue spectrum respect to secondaries ratio $c$.

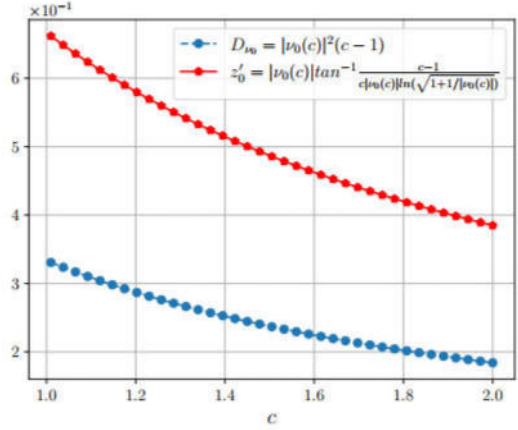

Figure C.12: Analytical diffusion coefficient and extrapolated distance trends with secondaries ratio $c$.

In Figure C.12, the trend of $D_{\nu_0}$ with the secondaries ratio is shown. For $c \approx 1$, the familiar value $1/3$ in a unit of the mean free path is recovered. This can be demonstrated by noticing that since $\nu_0 \to \infty$ in the proximity of $c \approx 1$, the inverse hyperbolic tangent in the root equation can be expanded in the Maclaurin series. Truncating the series to the first two terms and isolating the approximate expression of $\nu_0$, we get the expression of the diffusion coefficient. Moreover, by deriving Eq. (C.11) respect to $x$ and then using Eq. (C.12) we get

$$\frac{d^2 \Phi_0(x)}{dx^2} + \frac{1}{|\nu_0|^2} \Phi_0(x) = 0, \qquad (C.13)$$

which is the well-known diffusion equation, where $1/|\nu_0|^2$ plays the role of the square "buckling". If we refer to the commonly used no re-entrant current boundary condition of Eq. (9), we notice that by substituting the expression for the angular flux obtained in Eq. (C.8), with $A(\nu) = 0$, we can get an approximated expression for the critical length and extrapolated distance, indicated respectively as $a'$ and $z_0'$

$$\begin{aligned} a' &= \pi |\nu_0| - 2 z_0', \\ z_0' &= |\nu_0| tan^{-1} \frac{c-1}{c|\nu_0| ln(\sqrt{1+1/|\nu_0|})}. \end{aligned} \qquad (C.14)$$

For $c - 1 \ll 1$, it is easy to demonstrate that the extrapolated distance can be expanded to

$$z_0' \simeq \frac{1}{\sqrt{3(c-1)}} tan^{-1} 2 \left[ \frac{c-1}{3} \right]^{1/2}. \qquad (C.15)$$

The trend of $z_0'$ with the secondary ratio $c$ is also shown in Figure C.12. According to the definition of the extrapolated distance, $z_0'$ can be simply expressed as $\zeta_{\nu_0} D_{\nu_0}$, with all the quantities expressed in the unit of mean free path.

## Appendix D. Test specifications

**Table Appendix D.1**
1G cross section dataset of benchmark test cases. Cross sections in units of $cm^{-1}$.

| Material | $\nu$ | $\sigma_f$ | $\sigma_c$ | $\sigma_s$ | $\sigma_t$ | $c$ |
|---|---|---|---|---|---|---|
| Pu-239 (PUa) | 3.24 | 0.081600 | 0.019584 | 0.225216 | 0.32640 | 1.50 |
| Pu-239 (PUb) | 2.84 | 0.081600 | 0.019584 | 0.225216 | 0.32640 | 1.40 |
| H2O (refl) | 0.0 | 0.0 | 0.032640 | 0.293760 | 0.32640 | 0.90 |
| U-235(e) | 2.50 | 0.069227 | 0.010137 | 0.328042 | 0.40740 | 1.230 |
| Fe | 0.0 | 0.0 | 0.000465 | 0.232094 | 0.23256 | 0.9980 |
| Na | 0.0 | 0.0 | 0.0 | 0.086368 | 0.08636 | 1 |

**Table Appendix D.2**
2G cross section dataset of the benchmark test case. Cross sections in units of $cm^{-1}$.

| Material | $\nu_2$ | $\sigma_{2f}$ | $\sigma_{2c}$ | $\sigma_{22s}$ | $\sigma_{12s}$ | $\sigma_2$ | $\chi_2$ |
|---|---|---|---|---|---|---|---|
| PU | 3.24 | 0.081600 | 0.019584 | 0.225216 | 0.32640 | 1.50 | 0.575 |

| Material | $\nu_1$ | $\sigma_{1f}$ | $\sigma_{1c}$ | $\sigma_{11s}$ | $\sigma_{21s}$ | $\sigma_1$ | $\chi_1$ |
|---|---|---|---|---|---|---|---|
| PU | 0.0 | 0.0 | 0.032640 | 0.293760 | 0.32640 | 0.90 | 0.425 |

**Table Appendix D.3**
1G-2G homogeneous reference critical length $r_c$.

| Problem Id. | Geometry type | $r_c$ (mfp) | $r_c$ (cm) |
|---|---|---|---|
| PUb-1-0-SL | Slab | 0.736603 | 2.256751 |
| PU-2-0-SL | Slab | 0.396469 | 1.795602 |

**Table Appendix D.4**
1G heterogeneous reference critical lengths $r_c$.

| Problem Id. | Geometry type | $PU_a$ (mfp) | $PU_a$ (cm) | $H_2O$ (mfp) | $H_2O$ (cm) |
|---|---|---|---|---|---|
| PUa-H2O(0.5)-1-0-SL | Slab | 0.43015 | 1.317862 | 0.5 | 1.531863 |

**Table Appendix D.5**
1G nonsymmetric heterogeneous reference critical lengths $r_c$.

| Problem Id. | Geometry type | $U_{235}$ $(Ue)$ (mfp) | $U_{235}$ $(Ue)$ (cm) | $Fe$ (mfp) | $Fe$ (cm) | $Na$ (mfp) | $Na$ (cm) |
|---|---|---|---|---|---|---|---|
| PUa-H2O(0.5)-1-0-SL | Slab | 2.0858098 | 5.119720083 | 0.0738 | 0.317337461 | 0.173 | 2.002771002 |